\documentclass[twocolumn]{aastex631}
\usepackage{natbib,graphics,graphicx,multirow,amsmath,cancel,color,xcolor,hyperref,enumitem}
\usepackage{ulem}
\usepackage{bm}
\usepackage{float}
\usepackage{cleveref}
\usepackage[normalem]{ulem}
\definecolor{brightmagenta}{rgb}{1.0, 0.0, 0.7}

\crefname{equation}{Eq}{Eqs}
\crefrangelabelformat{equation}{(#3#1#4--#5#2#6)}
\usepackage{comment}

\shorttitle{}

\begin{document}
\title{Wave Particle Interaction in the Upstream of ICME Shocks}
\correspondingauthor{Shanwlee Sow Mondal}
\email{shanwlee.sowmondal@gmail.com\\
sowmondal@cua.edu
}

\author[0000-0003-4225-8520]{Shanwlee Sow Mondal}
\affil{Astronomy and Astrophysics Division, Physical Research Laboratory, Ahmedabad 380009, India}
\affil{The Catholic University of America, 620 Michigan Ave., N.E. Washington, DC 20064 USA}
\affil{NASA Goddard Space Flight Center, 8800 Greenbelt Rd., Greenbelt, MD 20771, USA}

\author[0000-0002-4781-5798]{Aveek Sarkar}
\affil{Astronomy and Astrophysics Division, Physical Research Laboratory, Ahmedabad 380009, India}

\author[0000-0002-2358-6628]{Sofiane Bourouaine}
\affil{Department of Aerospace, Physics and Space Sciences, Florida Institute of Technology, Melbourne, FL 32901}

\begin{abstract}

Shocks associated with Interplanetary Coronal Mass Ejections are known to energize charged particles and give rise to Solar Energetic Particles. Many of these energetic particles move ahead of the shock to create a foreshock region. The foreshock region primarily consists of solar wind plasma, exhibiting turbulent velocity and magnetic fields. Such turbulent behavior results from inherent solar wind turbulence modified by energetic particles. We analyze magnetic field data from six such ICME shocks observed by the Wind spacecraft. The analysis of the shock upstream shows that the magnetic power spectral density (PSD) maintains a power-law slope of $-5/3$. We also identify clear intermittent peaks in the PSD. After characterizing these peaks, we investigate various possibilities for their generation. Our analysis indicates that these peaks in the PSD are due to the resonant interaction of Alfv\'en waves with the bulk solar wind protons and protons with energy up to $10$~keV. However, evidence of Alfv\'en wave interaction with highly energetic protons is not evident in our analysis, and we anticipate that such evidence is obscured by the prevailing solar wind turbulence in the shock upstream.

\end{abstract}


\section{Introduction}\label{sec:introduction}
In situ observations of Interplanetary Coronal Mass Ejections (ICMEs) reveal that the shocks ahead of ICMEs are one of the primary sources of Solar Energetic Particles (SEPs). Diffusive Shock Acceleration (DSA) and Shock Drift Acceleration (SDA) are the most prominent mechanisms thought to accelerate charged particles. The dominant mechanism depends on the angle between the shock normal and the upstream magnetic field direction. A comprehensive numerical study of these acceleration processes is conducted in ~\cite{Shanwlee_2021}; which also proposes multiple observational signatures as a result of these processes.

It is envisaged that charged particles once accelerated at the shock front, escape and travel further upstream, influencing the ambient solar wind plasma. These energetic particles disturb the upstream region upon escape, leading to Alfv\'enic fluctuations in the magnetic field. There is a possibility that these charged particles may interact with these fluctuations and, if necessary conditions are satisfied, may give rise to resonant or non-resonant instabilities~\citep{Shanwlee_2021}. The signature of such instabilities should be reflected in the power spectrum of the shock upstream data. Understanding heliospheric shocks may also provide insights into the long-standing astrophysical problem of particle acceleration in shocks.

Several efforts, both in theory and observation, have been made to understand the waves and instabilities in and around ICME shocks. \cite{Lee_83} proposed a quasi-linear theory describing the excitation of hydromagnetic waves and particle acceleration in interplanetary shocks. Using magnetic field data from the Advanced Composition Explorer (ACE) spacecraft, \cite{Bamert_2004} studied the wave excitation in the upstream region of an interplanetary shock. They concluded that in addition to the quasi-linear theory proposed by \cite{Lee_83}, further understanding is required to explain all wave-particle interactions occurring in the shock upstream.

\cite{Vinas_1984JGR....89.3762V} observed the association of proton beams and MHD waves in the vicinity of interplanetary shocks. They observed that the magnetic power spectra near these shocks are significantly steeper at high frequencies compared to those in the ambient solar wind. Their analysis indicates that the interaction between the proton beam and waves leads to either resonant or non-resonant electromagnetic ion beam instability.

~\cite{Wilson_2009} found evidence of low-frequency waves ($0.25$ Hz $< f < 10$ Hz) at the upstream of five quasi-perpendicular interplanetary (IP) shocks. Their analysis revealed characteristics of these waves that are mostly consistent with whistler waves. While studying several interplanetary shocks observed by the STEREO spacecraft, \cite{Blanco-Cano_2016JGRA..121..992B} found that Whistler waves and Ultra Low Frequency (ULF) waves can permeate the shock upstream region. In the case of quasi-parallel shocks in the upstream region, these waves interact with ions that are reflected by the shock or escaping into the upstream region, generating ion instabilities. Many other authors also have reported the presence of whistler waves in IP shocks~\cite{Wilson_2012, Wilson_2017, Kajdic_2012, Ramirez_2012}. 
In case of quasi perpendicular shock, \cite{Zank_2006JGRA..111.6108Z} reported no evidence of enhanced wave activity or turbulence in the shock upstream region although particles are still accelerated.

While studying wave activities in both upstream and downstream of the interplanetary (IP) shock,~\cite{Kajdic_2012} found evidence indicating the presence of upstream wave activities and its linkage with the increase in suprathermal ions having energies less than $20$ keV. 

Using data from the ACE spacecraft and two ARTEMIS satellites, \cite{Kajdic_2017} conducted a study on an interplanetary shock event. They discovered evidence of varied suprathermal particle distributions in the shock upstream. Furthermore, they observed changes in shock properties, such as the Mach number, across the shock surface, and noted corresponding variations in the suprathermal particle distribution in the upstream region.

In a recent study,~\cite{Zhao_2021} investigated turbulence within an ICME shock, simultaneously observed by the Solar Orbiter spacecraft at 0.8 AU and the Wind spacecraft at 1 AU. The observations revealed an enhancement across a broad low-frequency band. Although the authors did not conclusively identify the nature of these upstream waves, they suggested the possible excitation of MHD waves by streaming particles. 

Interplanetary shock waves have the ability to alter the turbulence characteristics of the upstream solar wind. The shock serves a dual role, as outlined by \citep{Zhao_2021} -- it can excite unstable modes due to the streaming of high-energy particles, while, at the same time, it can amplify the pre-existing solar wind turbulence. The finite difference in kinetic energy between the upstream and downstream flows acts as a free energy source in the shock wave, which must be dissipated to conserve energy. This free energy can either be utilized to scatter the particles by exciting turbulence in the shock vicinity or can be dissipated by wave-particle interaction resulting in the observed steepening of the magnetic spectrum in the dissipation range.

Primarily using data from the ACE spacecraft, \cite{Perri_2023ApJ...950...62P} established a connection between magnetic field fluctuations and energetic particles in three interplanetary shocks. Their results also show that magnetic field fluctuations upstream of the shock contribute to particle acceleration, with more energetic particles escaping the shock front and propagating ahead.

The contemporary understanding of upstream plasma turbulence and wave-particle interaction in the foreshock region has been discussed by~\cite{Pitvna_2021}. Earlier, while investigating the Bastille Day CME, \cite{Bamert_2008} found that the turbulence generated by energetic particles in the shock upstream exhibits a magnetic spectrum with a power-law slope of $-3/2$.

Before delving deeper into the topic, it is worth noting that the radial interplanetary magnetic field (IMF) often exhibits wave-like behavior. Using fifteen years of data from the Wind spacecraft \cite{Pi_atmos13020173}, examined magnetic field fluctuations across various radial intervals when the IMF aligns with the solar wind's flow velocity. They observed frequent wavy structures with frequencies above 0.01 Hz, likely driven by proton cyclotron instability. However, it's important to note that multiple types of waves and instabilities may be present in this region.

Here, we investigate the upstream regions of six ICME shocks observed by the Wind spacecraft. All the shocks analyzed in this study show enhancement in particle fluxes over a wide energy range, well before the shock arrival. 
 
Our main objective is to investigate the wave activity in the shock upstream region and its possible connection to the energetic particles present in the surrounding medium. While most of the time, the power spectrum maintains a typical Kolmogorov-like spectrum with a power law slope of $-5/3$, we occasionally encounter a distinctive peak in the power spectrum. We attribute these peaks to the interaction between the particles and the waves. Additionally, we investigate the conditions under which such power spectrum peaks can be detected.

The rest of the paper is structured as follows: Section \ref{Section:Observation} describes the properties of the ICME-driven shocks chosen for this study, along with the particle data from Wind. Section \ref{Section:Data analysis} outlines how magnetic fluctuations in the shock upstream are analyzed and waves are detected. In Section \ref{Section:Discussion}, we discuss our analysis results and explore the possible mechanisms of wave excitation. Finally, Section \ref{Section:Summary} provides a summary of our findings.

\section{Observations}\label{Section:Observation}

\begin{table*}
\centering
 \begin{tabular}{c c c c c c c} 
\hline
 Shock & Shock arrival  &  $\theta_{BN}$ & $v_{A}$       & $v_{SW}$      & $v_{sh}$      & $M_{A}$ \\
       &    time        &                & (km s$^{-1}$) & (km s$^{-1}$) & (km s$^{-1}$) &    \\[1ex] 
\hline
\multirow{2}{*}{1} & $2001/09/25$ & \multirow{2}{*}{$82^\circ$} & \multirow{2}{*}{$82$} &\multirow{2}{*}{$340$} &\multirow{2}{*}{$573$} &\multirow{2}{*}{$5.2$} \\
                   &$20:17:18$    &            &       &       &       &      \\
 \hline
\multirow{2}{*}{2} & $1998/01/24$ & \multirow{2}{*}{$64^\circ$} & \multirow{2}{*}{$19$} &\multirow{2}{*}{$370$} &\multirow{2}{*}{$356$} &\multirow{2}{*}{$4.1$} \\
                   &$04:37:30$    &            &       &       &       &      \\
                   \hline
\multirow{2}{*}{3} & $2013/04/13$ & \multirow{2}{*}{$40^\circ$} & \multirow{2}{*}{$45$} &\multirow{2}{*}{$367$} &\multirow{2}{*}{$479$} &\multirow{2}{*}{$4.7$} \\
                   &$22:13:15$    &            &       &       &       &      \\
                  \hline

\multirow{2}{*}{4} & $1995/11/27$ & \multirow{2}{*}{$62^\circ$} & \multirow{2}{*}{$27$} &\multirow{2}{*}{$306$} &\multirow{2}{*}{$327$} &\multirow{2}{*}{$2.8$} \\
                   &$08:21:51$    &            &       &       &       &      \\
                   \hline
\multirow{2}{*}{5} & $1997/05/15$ & \multirow{2}{*}{$86^\circ$} & \multirow{2}{*}{$37$} &\multirow{2}{*}{$324$} &\multirow{2}{*}{$438$} &\multirow{2}{*}{$8.7$} \\
                   &$01:15:2$    &            &       &       &       &      \\
                   \hline
\multirow{2}{*}{6} & $1997/10/10$ & \multirow{2}{*}{$89^\circ$} & \multirow{2}{*}{$52$} &\multirow{2}{*}{$400$} &\multirow{2}{*}{$448$} &\multirow{2}{*}{$1.7$} \\
                   &$15:57:06$    &            &       &       &       &      \\
 \hline
 \end{tabular}
 \label{tab:table_1}
 \caption{List of shocks analyzed in this work. All were detected by the Wind spacecraft. $\theta_{BN}$ is the angle between the shock propagation direction and the upstream magnetic field. $V_{A}$, $v_{SW}$, and $v_{sh}$ are the Alfv\'en speed, solar wind speed, and shock speed, respectively. $M_A$ stands for the Alfv\'enic mach number. Shock parameters are taken from the Heliospheric Shock database, developed and maintained by the University of Helsinki [\url{http://ipshocks.fi}].}
\end{table*}

To investigate the nature of waves in the upstream region of ICME shocks we utilize high-resolution magnetic field data ($0.092$s cadence) from the Wind/Magnetic Field Instrument \citep[Wind/MFI;][]{Lepping_1995}, solar wind plasma data from the Wind/Solar Wind Experiment \citep[Wind/SWE;][]{Ogilvie_1995}, and particle data from the Wind/3DP \citep{Lin_1995} instrument. All these data are obtained from the NASA Goddard Space Flight Center Coordinated Data Analysis Web [\url{http://cdaweb.gsfc.nasa.gov/}]. The analyzed ICME shocks are listed in Table \ref{tab:table_1}. Shock parameters in the table are obtained from the shock database [\url{http://ipshocks.fi}].

\textit{Shock selection criteria} --  All the selected shocks are foreshock events and are detected by the Wind spacecraft, with the Alfv\'enic Mach reasonably high, so that they are capable of accelerating particles. Preference is given to shocks exhibiting a high energy particle flux enhancement at the time of the shock detection. The shocks chosen for this study are all clear shock events, i.e., none of them are followed by another shock - at least not in the three days leading up to or following the arrival of any of the shocks under consideration. This was done in order to ensure that the results are only influenced by the shock that was detected and not by contamination from additional shock events. 

Firstly, we provide a detailed analysis of one of the shocks (Shock 2 from Table \ref{tab:table_1}), and then we present an analysis of the remaining shocks.

\begin{figure*}[htbp!]
  \centering
      \includegraphics[width=0.9\textwidth,angle=0]{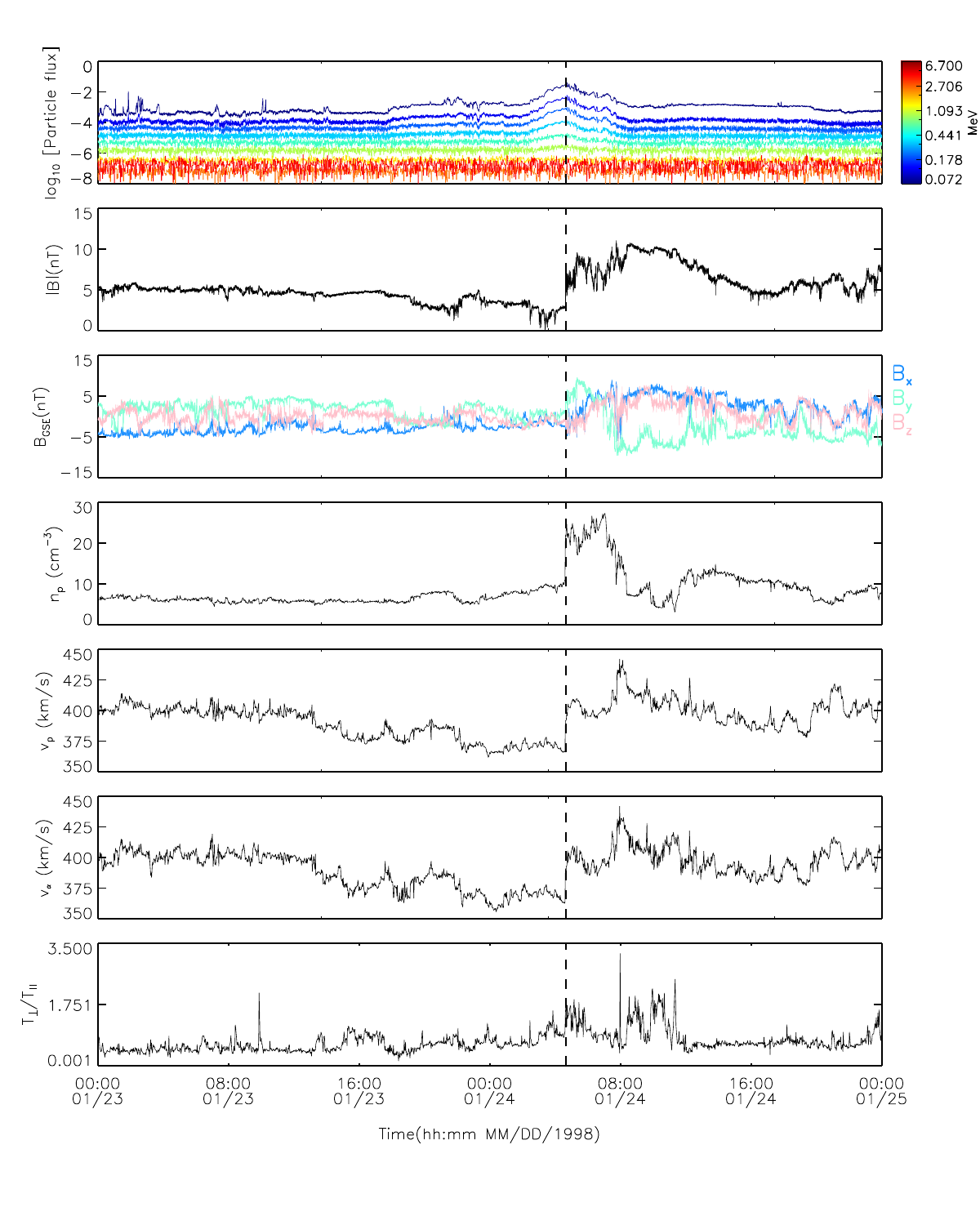}
         \caption{A typical shock chosen from Table \ref{tab:table_1}, identified as shock 2. The vertical dashed line denotes the arrival time of the shock. The upper plot displays the particle flux of various energetic particles (from $72$ keV to $6.7$ MeV) near the shock occurrence. Parameters such as the magnitude of the magnetic field ($|B|$), the three components of the magnetic field ($B_x$, $B_y$, $B_z$), particle density ($n_p$), bulk proton velocity ($V_p$), bulk $\alpha$ particle velocity ($V_{\alpha}$), and the evolution of temperature anisotropy ($T_{\perp}/T_{\parallel}$) are graphically represented around the time of the shock arrival.}
  \label{case1_shock}
  \end{figure*}

Figure \ref{case1_shock} shows the solar wind plasma parameters close to the shock front. The Wind spacecraft detected the shock on January $24$, $1998$ at $04$:$37$:$30$ UT and is marked by the red dashed line. Simultaneous particle flux measurements are shown in figure \ref{case1_shock} which shows enhancement in particle fluxes well before the shock arrival. Particles with energies $72$ keV to $6.7$ MeV are observed to form a foreshock region. Apart from that, particles with energies $0.4-3$ keV and $72$ keV to $6.7$ MeV also show local enhancements in the far upstream region. The presence of such particle enhancements, both close and away from the shock, can locally excite unstable modes and can only be identified if the power of the unstable mode is well above the background solar wind turbulence. 

\section{Data analysis}\label{Section:Data analysis}
To understand the nature of unstable modes in the upstream region of the ICME shock, we generated magnetic power spectra at different locations in the shock upstream region. Each spectrum is derived within a time window of width $5, 15, 45$ and $60$ minutes. Windows with different widths aid in identifying any unusual power enhancements that might remain unnoticed by larger windows, where the dominant nature of the background turbulence is more expected. Each time window was then moved through the upstream region, from the far upstream to the shock front, generating magnetic power spectral density (PSD) at each consecutive time window (with intervals determined by the selected window width). In this study, we discuss the results corresponding to a time window of width $5$ minutes.

For each time window placed at any given upstream location, the total power spectral density corresponds to the sum of individual power spectral density obtained for each magnetic field component and is given by the following,

\begin{equation*}
  PSD = PSD_{x} + PSD_y + PSD_z
\end{equation*}

where $PSD_i$ corresponds to the power spectrum of $(Bi-B_{i0})$ with $i=x, y$ and $z$. $B_{i0}$ is the mean value of $B_{i}$ in the corresponding 5 minutes window. We have used \textit{fft\textunderscore powerspectrum} routine in IDL to estimate the power spectrum of individual magnetic field components.

\begin{figure}[h]
  \centering
      \includegraphics[width=0.5\textwidth,angle=0]{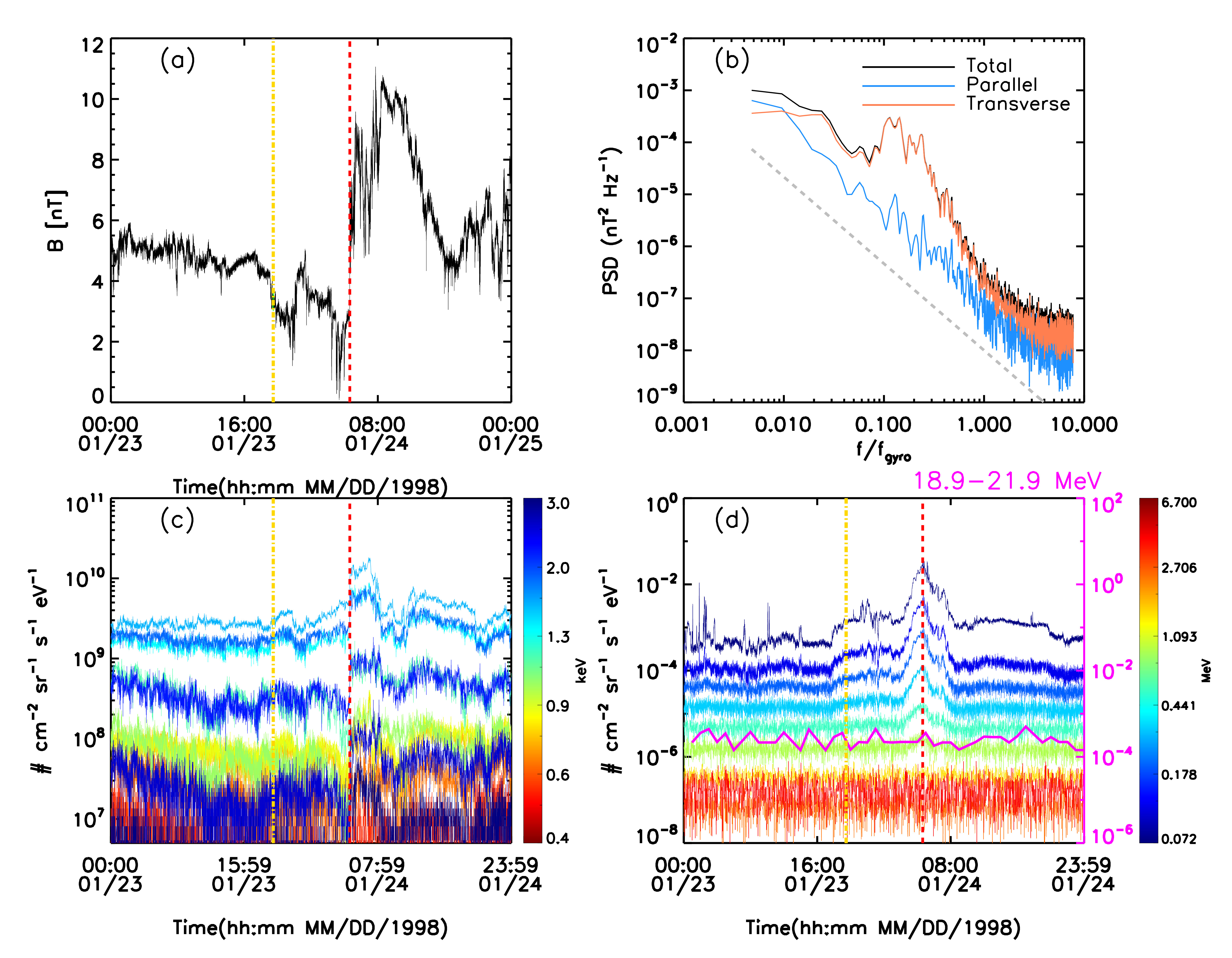}
 \caption{Parameters depicting the presence of unstable modes in the shock upstream. The vertical red dashed lines in each plot indicate the location of the shock. (a) The magnitude of the magnetic field around the shock. (b) Magnetic power spectral density (PSD) showing the peak, attributed to the transverse component of the magnetic field. PSD is computed over a $5$-minute time window at the time marked with the yellow vertical line in the time series plots. The gray dotted line shows the Kolmogorov spectrum with slope $-5/3$. (c) Time series data of the low energy particles ($0.4-3$ keV). (d) Time series data of the highly energetic particles ($72$ keV$-21.9$ MeV).}
  \label{psd_1}
  \end{figure}

A careful examination of the series of power spectral densities reveals intermittent peaks in the power spectrum. However, given the need to analyze a large number of PSDs, we have developed an automated peak detection algorithm, the details of which is provided in the Appendix~\ref{section:Peak Detection Algorithm}. It is worth mentioning that such peaks in the shock upstream have been observed earlier as well \citep{Zhao_2021}.  We decompose the fluctuations in the magnetic field into components parallel and transverse to the local mean magnetic field in order to obtain a better understanding of the directional nature of the dominant power. Our study indicates that the transverse component dominates the PSD peaks. An illustration of our findings is presented in Figure~\ref{psd_1}.

A strong peak in the power spectral density is detected around a frequency of $\sim 0.1 f_{gyro}$ in Figure~\ref{psd_1}(b), where $f_{gyro}$ is the gyro-frequency calculated using the local plasma parameters at the time marked by the yellow dashed line within a 5-minute time window. The gray dotted line shows the Kolmogorov spectrum with slope $-5/3$. Plots ~\ref{psd_1}(c) and ~\ref{psd_1}(d) demonstrate the evolution of particle fluxes in energy range $0.4-3$ keV and $72$ keV to $6.7$ MeV, respectively. The magenta line in plot ~\ref{psd_1}(d) corresponds to particle fluxes in energy range $18.9-21.9$ MeV.

Time intervals where PSD peaks are detected for the all the shocks analyzed in this study are shown in the Appendix~\ref{section:SD}.

 \section{Analysis and Discussion}\label{Section:Discussion}
 Most of the shocks analyzed here demonstrate an enhancement of suprathermal particle flux even before the shock arrival (e.g., see Figure~\ref{case1_shock}). Hereafter we discuss the possible sources of free energy and the background solar wind condition responsible for the observed peaks in the magnetic power spectrum.
 
\subsection{Proton-alpha drift velocity}
\begin{figure}[h]
  \centering
     \includegraphics[width=0.5\textwidth,angle=0]{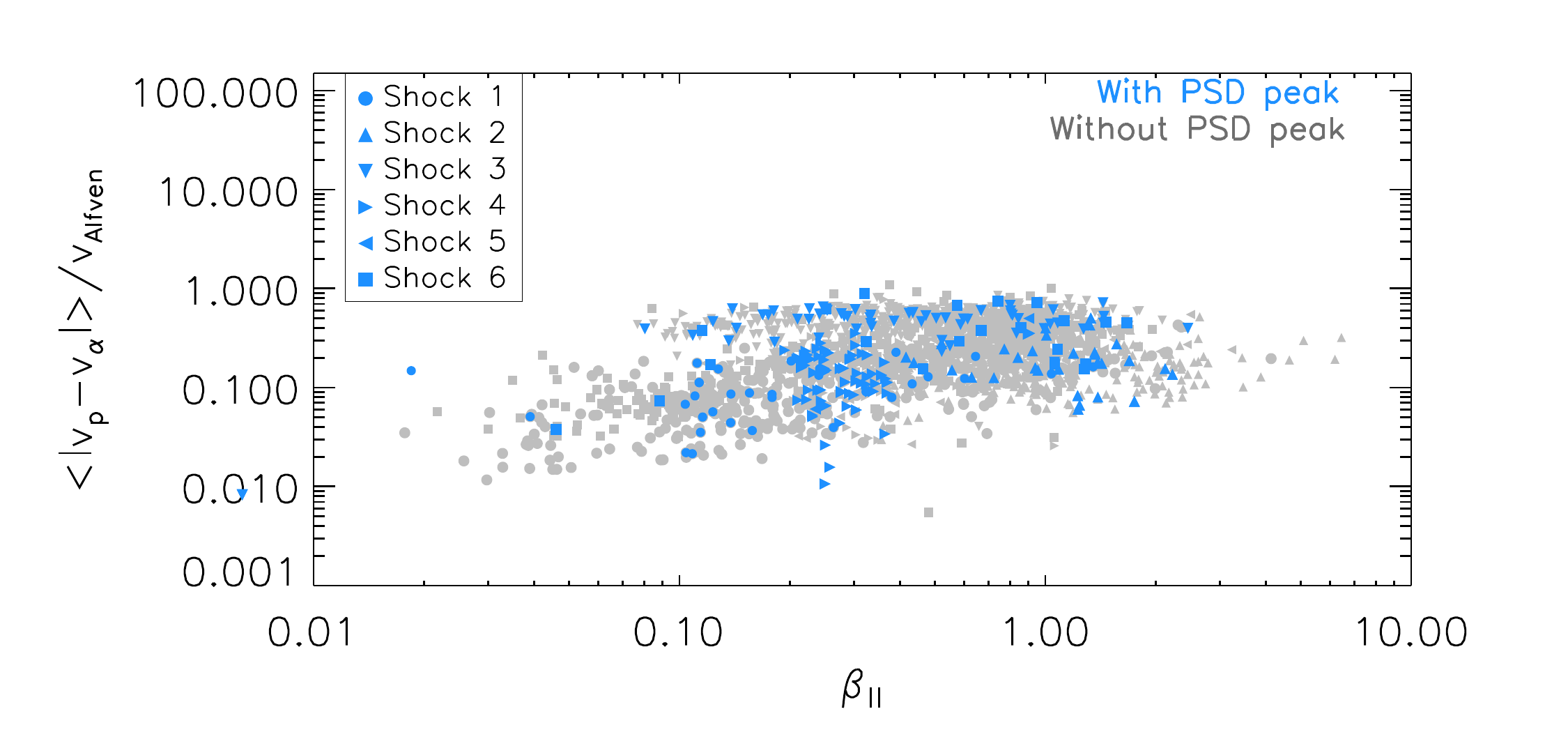}
       \caption{The plot suggests that the peaks in the power spectra are not caused by the instability created by the $\alpha$-proton velocity difference. The information from the power spectra of all shocks is accumulated in this plot, where only the points marked in blue indicate power spectra peaks}
  \label{fig:thermal_check}
  \end{figure}

Figure \ref{fig:thermal_check} shows the relative drift speeds between proton and alpha particles in comparison to the local Alfv\'en speed. As previously mentioned, we conducted this analysis by employing a sliding $5$-minute time window across the shock upstream. Each data point corresponds to the results obtained from a single $5$-minute time window. Points marked in blue indicate intervals with PSD enhancement, while the remaining intervals without PSD peak/enhancement are marked in gray. Notably, the detected PSD peaks are consistently associated with a relative drift that is much less than the local Alfv\'en speed or, at most, equal to the Alfv\'en speed. Thus, it is possible that the free energy in the alpha-proton drift speeds is not sufficient to generate instabilities in the observed solar wind plasma, which rules out it as the source of the PSD peaks.

\subsection{Proton temperature anisotropy} 
\begin{figure}[h]
  \centering
      \includegraphics[width=0.5\textwidth,angle=0]{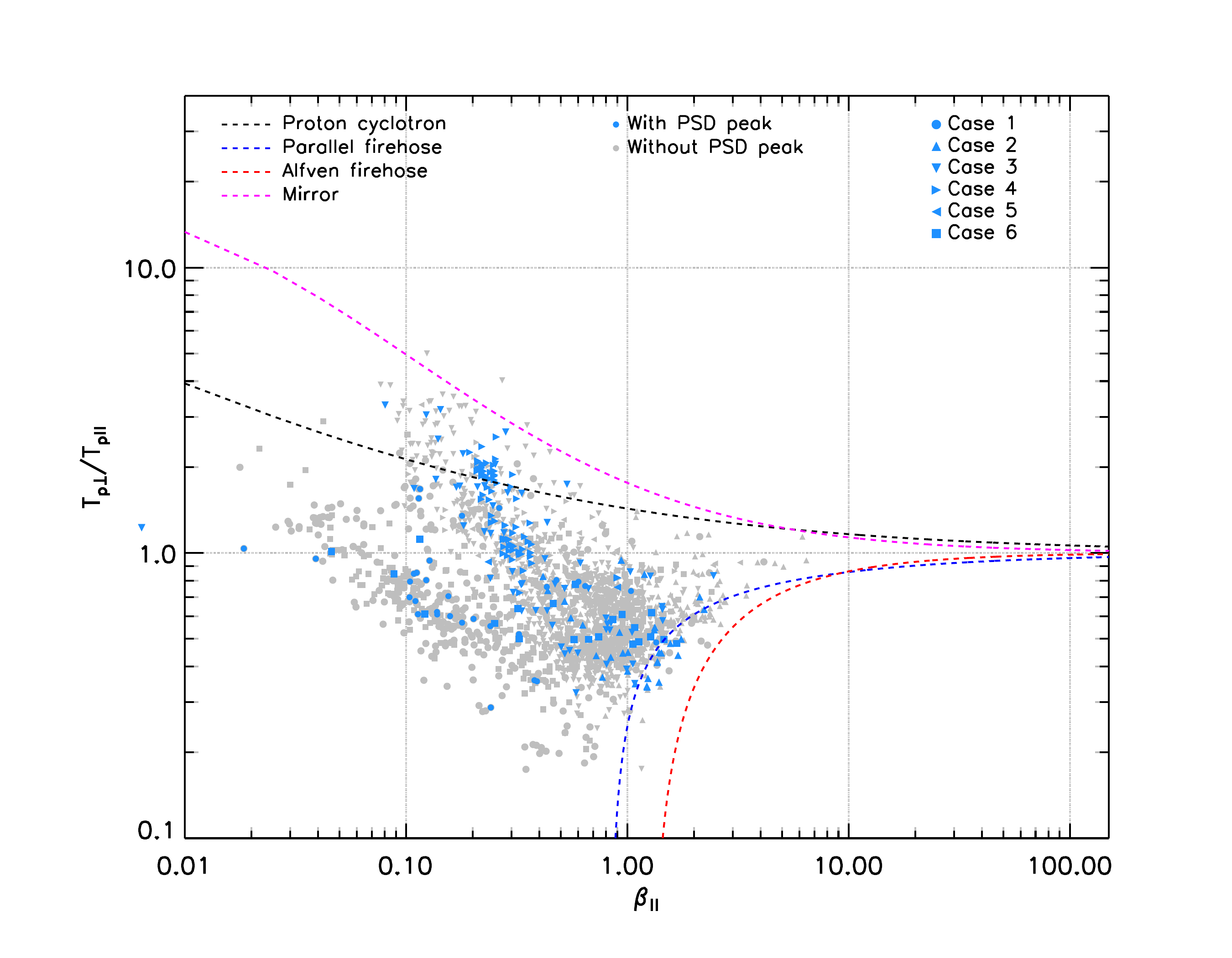}
       \caption{Scatter plot illustrating the data distribution in the proton temperature anisotropy ratio and plasma $\beta_{||}$ plane. Data points with and without PSD peaks are indicated by grey and blue points, respectively. The overlaid curves are thresholds for the proton cyclotron (black), parallel firehose (blue), Alfv\'en firehose (red), and mirror (magenta) instabilities. As shown in the plot, some of the data points with PSD peaks hint towards possible activation of proton cyclotron and parallel firehose instabilities.}
  \label{fig:mirror_modes}
  \end{figure}

Next, we check whether instability driven by the proton-temperature anisotropy is responsible for the observed PSD peaks. Figure~\ref{fig:mirror_modes} shows the proton temperature anisotropy ratio ($T_{p \perp}/T_{p \parallel}$) analyzed for each $5$ minute window in the upstream region of shock 2. Once again, points with PSD peaks are highlighted in blue, whereas those without PSD peaks are represented in gray. The overlaid curves indicate thresholds for proton cyclotron, parallel firehose, Alfv\'en firehose, and mirror instability limits, obtained from \cite{Hellinger_2006} for a growth rate of $\gamma = 10^{-3} \Omega$, where $\Omega$ is the proton cyclotron frequency. The relation used to derive the curves are following equation~\ref{tempaniso}.

\begin{equation}\label{tempaniso}
    \frac{T_{p\bot}}{T_{p\parallel}} = 1 + \frac{a}{(\beta_{\parallel p} - \beta_{0})^b}
\end{equation}
where the $a$, $b$ and $\beta_{0}$ coefficients for different instabilities are given by the table below.

\begin{table}[htbp!]
    \centering
    \begin{tabular}{|c|c|c|c|}
    \hline
     Instability & a & b & $\beta_{0}$ \\
     \hline
     
       Proton cyclotron   & 0.43 & 0.42 & -0.0004 \\
       Parallel firehose  & -0.47 & 0.53 & 0.59 \\
       Alfvén firehose    & -1.4 & 1.0 & -0.11 \\
       Mirror             & 0.77 & 0.76 & -0.016 \\
       \hline
    \end{tabular}
    \caption{Instability threshold parameters for $\gamma = 10^{-3} \Omega$ from \cite{Hellinger_2006}.}
    \label{tab:my_label}
\end{table}

Although most of the temperature ratios corresponding to the PSD peaks are within the stable region in the parameter space, some of them have crossed the proton cyclotron and parallel firehose instability thresholds. Thus, the aforementioned temperature anisotropy instabilities have the potential to cause some of the PSD peaks. 

\subsection{Particle streaming instability }
Furthermore, we explore the possibility of streaming instability induced by solar wind particles. A particle can resonantly interact with a left-hand polarized wave having a frequency similar to the particle's gyro-frequency. This resonance interaction facilitates energy exchange between the wave and the particle, manifesting as peaks or enhancements in the PSD plots over a certain frequency range.

Frequencies corresponding to different length scales are Doppler shifted to the spacecraft frame through the relation
\begin{equation}
    f_{sc} = f_{sw} + \frac{\bm{k}\cdot\bm{V}_{sw}}{2\pi}  
\end{equation}

Here, $f_{sc}$ and $f_{sw}$ are the wave frequencies in the spacecraft and plasma frame, respectively. $\bm{k}$ is the wave vector and $\bm{V}_{sw}$ is the solar wind velocity.

In all the cases discussed in this work, the Alfv\'en speed is much smaller than the solar wind speed. Hence we can write the above equation as, \\

\begin{equation}\label{costheta}
    f_{sc} \approx \frac{\bm{k} \cdot \bm{V}_{sw}}{2\pi} = \frac{kV_{sw}}{2\pi} cos\theta
\end{equation}

Here, $\theta$ is the angle between the wave vector $\bm{k}$ and the solar wind velocity $\bm{V}_{sw}$.

Considering the frequency corresponding to the PSD peak to be the Doppler-shifted gyrofrequency of a particle resonating with the wave in the spacecraft frame, one can estimate the energy of this resonating particle. If $f_{\text{peak}}$ is the frequency (in the spacecraft frame) at which the power spectrum shows enhancement or peaks, and $k$ is the wave number in the plasma frame, following Equation ~\ref{costheta} one can write 

\begin{equation}
    f_{peak} \approx \frac{kV_{sw}}{2\pi}cos\theta 
\end{equation}

In order to undergo resonant interaction, the particle's gyro-radius should be comparable to the wave length of the resonating wave. Thus $k \approx \frac{2\pi}{r_{g}}$ where, $r_{g} (=\frac{mv_{\perp}}{qB})$ corresponds to the gyroradius of the particle in resonance with the wave, $q$ is the charge of the particle and $B$ is the background magnetic field. Once we know the corresponding velocity ($v_{\perp}$) in plasma frame, we can convert it to the spacecraft frame ($v_{sc} \approx v_{\perp} + V_{sw}$) to know the energy of the particle in resonance with the wave.

\begin{figure}[h]
  \centering
      \includegraphics[width=0.525\textwidth,angle=0]{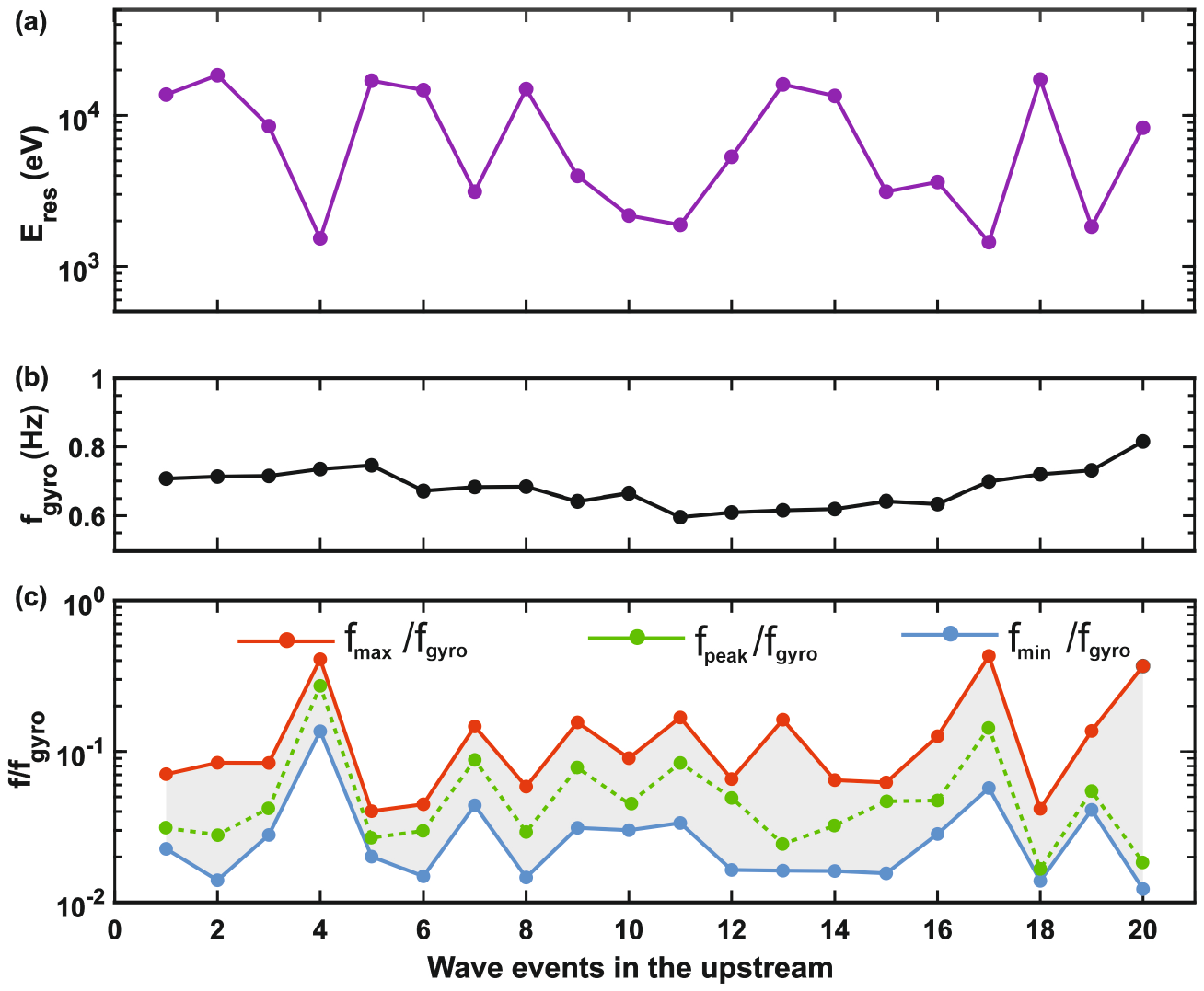}
        \caption{The plot illustrates the frequency ranges corresponding to wave activity at various upstream locations of shock 2. For any identified wave activity interval, the green dot indicates the frequency related to the PSD peak, while the blue and red dots mark the lower and upper frequencies encompassing the complete PSD-enhanced frequency range. All the frequencies are normalized with respect to the local gyro frequency ($f_{gyro}$) in spacecraft frame, indicated in the middle. At the top, the plot displays the corresponding particle energies capable of resonating with the waves at the frequencies of the PSD peaks.}
  \label{barplot_with_energy_case2}
  \end{figure}

Figure \ref{barplot_with_energy_case2} demonstrates the resonating frequency and particle energy corresponding to wave activity at various upstream intervals of shock 2. For this specific shock, a total of twenty intervals have been identified to be associated with PSD enhancement. In every interval, a green dot represents the frequency corresponding to the PSD peak, while blue and red dots, respectively, indicate the lower and upper frequencies covering the complete PSD enhanced range. All frequencies are normalized with the local gyro-frequency ($f_{gyro}$) in spacecraft frame and is shown by the black curve in the middle of the figure. The top magenta curve demonstrates the anticipated energy of the particle participating in the wave-particle resonance interaction. For this specific shock, the peak frequencies closely coincide with the gyro-frequencies of bulk solar wind protons or mildly energized suprathermal protons. This indicates that particles involved in these resonant interactions have energies mostly in the range $1 - 10$ keV. 

Nevertheless, we hardly ever detect evidence of PSD peaks connected to the high energy particle resonance interaction. One possible reason could be the dominance of background solar wind turbulence over the instabilities driven by higher-energy particles. Resonant interaction with high energy particles requires a longer wave length, which in turn corresponds to a smaller wave number and a smaller resonance frequency (from equation \ref{costheta}). Unless the instabilities induced by high energy particles are extremely powerful, the background solar wind turbulence predominates at lower frequencies, making identification of high energy particle driven unstable modes unlikely. Hence, those unstable modes do not appear in the PSD plots.

\subsection{Background solar wind condition and PSD peaks}
To gain a deeper understanding of the background magnetic fluctuation and the conditions for the PSD peak to become detectable, we calculate the mean square magnetic fluctuations ($\overline{\delta B^{2}}$) associated with each five-minute time window. Here,
\begin{equation}
    \overline{\delta B^{2}} = \frac{1}{N}\sum_{j=1}^{N}\sum_{i=1}^{3}(B_{i,j} - \bar{B}_{i,j})^2
\end{equation}

Next we determine the ratio $\frac{\overline{\delta B^{2}}}{\overline{B^{2}}}$, where
 
\begin{equation}
 \overline{B^{2}} = \frac{1}{N}\sum_{j=1}^{N}B_{j}^{2}
\end{equation}

represents the mean square background magnetic field over the five-minute time window. Here, $N$ is the number of data points within the five-minute interval, and the summation index $i = 1,2,3$ corresponds to $x,y,z$ components of the total magnetic field ($B$).

\begin{figure}[h]
  \centering
     \includegraphics[width=0.47\textwidth,angle=0]{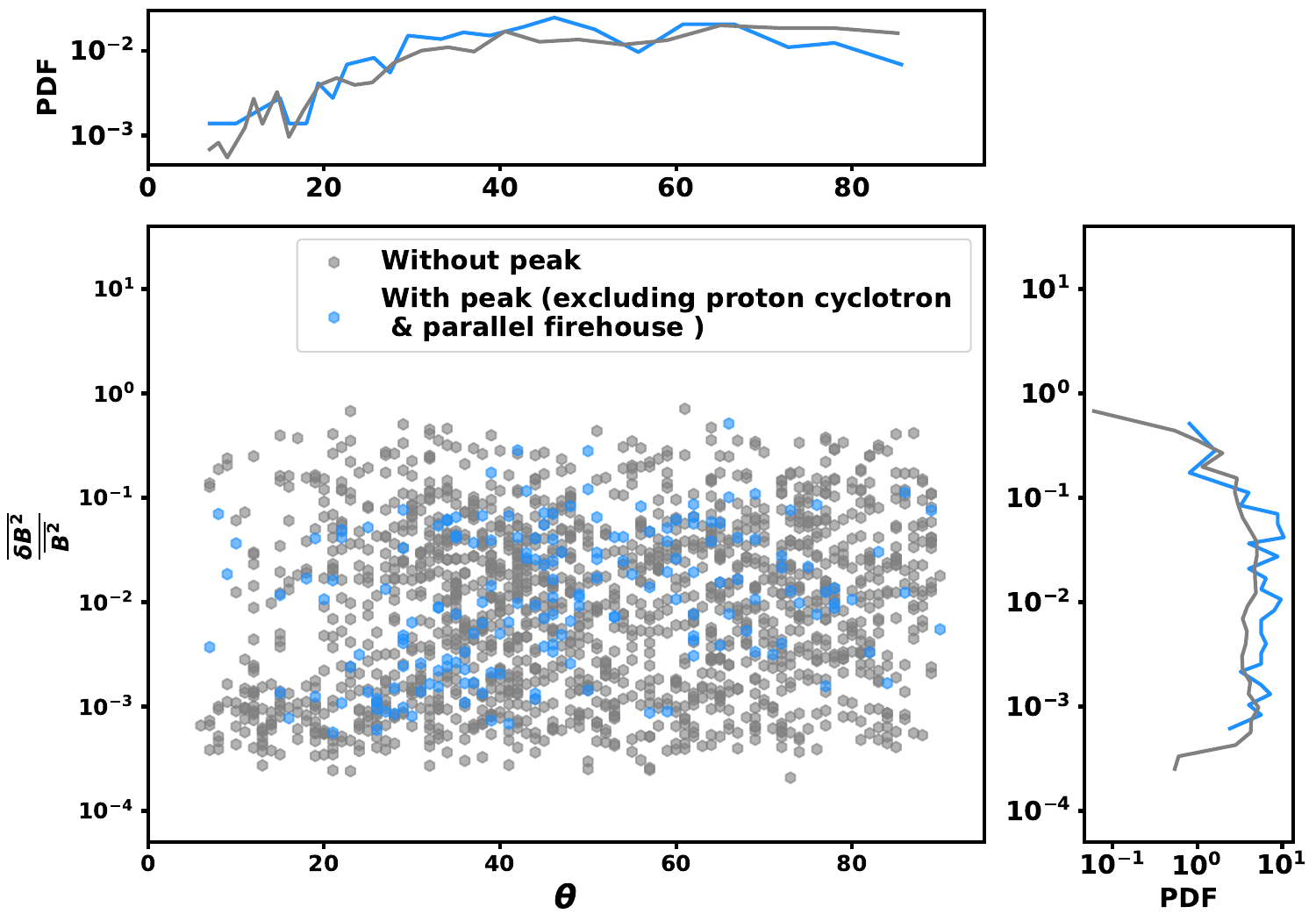}
       \caption{A scatter plot showing the mean square fluctuations ($\frac{\overline{\delta B^{2}}}{\bar{B^{2}}}$) associated with each five-minute window for all the shocks. The values are plotted against the angle $\theta$, representing the angle between the solar wind velocity and the mean magnetic field. Time windows corresponding to PSD peaks are marked with blue points, while time windows with no peaks are marked with gray points. PSD peaks associated with proton- cyclotron and parallel-firehose instabilities are excluded. The right panel of the scatter plot, aligned vertically along the axis, shows the PDFs of the gray and blue points in the $\frac{\overline{\delta B^{2}}}{\bar{B^{2}}}$ space, while the top panel displays the PDFs of the points in the $\theta$ space. Although the PDFs overlap, the KS test confirms that the gray and blue distributions are statistically distinct.} 
  \label{fig:statistics}
  \end{figure}

Besides the magnetic fluctuations, we also examine the angle ($\theta$) between the solar wind velocity and the magnetic field in each time window. We then produce a scatter plot of these data points in the $\frac{\overline{\delta B^{2}}}{\overline{B^{2}}}$ -- $\theta$ parameter space, as shown in Figure~\ref{fig:statistics}. In this plot, gray points represent time intervals where PSD peaks are not detected, while blue points correspond to intervals where PSD peaks are identified. Blue points that exceed the proton-cyclotron and parallel-firehose instability thresholds (visible in Figure \ref{fig:mirror_modes}) are excluded from the plot. The appearance of these blue and gray points may suggest a similar distribution; in other words, both sets of points appear to be nearly evenly distributed across the parameter space, suggesting their similar behavior. For a more qualitative analysis, we calculate the Probability Density Function (PDF) for the blue and gray points individually in $\frac{\overline{\delta B^{2}}}{\overline{B^{2}}}$ and $\theta$. The PDFs are obtained by computing histograms with logarithmic binning and normalizing the counts by the total area under the histogram. The PDF for $\theta$ is shown at the top of the scatter plot, while the PDF for $\frac{\overline{\delta B^{2}}}{\overline{B^{2}}}$ is plotted vertically along its own axis in Figure~\ref{fig:statistics}. The substantial overlap between these PDFs highlights the challenge of distinguishing PSD peaks due to the dominance of background turbulence. 

To determine whether the gray and blue distributions are statistically distinct, we perform a two-sample Kolmogorov-Smirnov (KS) test on these distributions with the null hypothesis that \textit{both blue and gray populations originate from the same parent distribution}. For the $\theta$ values of the gray and blue populations, the KS statistic is calculated as $0.1216$, with a corresponding $p$-value of $0.0165$.  Similarly, for the gray and blue populations of $\frac{\overline{\delta B^{2}}}{\overline{B^{2}}}$, the KS statistic is found to be $0.1155$ with a $p$-value of $0.0264$. Since both $p$-values are below $0.05$, this indicates that the gray and blue distributions are statistically different and likely originate from different underlying populations. Therefore,we can conclude that the background solar wind conditions in the $\frac{\overline{\delta B^{2}}}{\overline{B^{2}}}$ -- $\theta$ parameter space differ from those when the PSD peaks are detected.

It may be noted that the percentage of intervals with PSD peaks is very small in comparison to the total number of analyzed intervals. There are several factors that could account for this small fraction.

\textit{Firstly}, because of the background turbulence already present in the solar wind plasma, it is not easy to discern the waves caused by energetic particles. Any unstable modes in the system are often suppressed by the intrinsic turbulence of the solar wind, particularly in the lower frequency regime where the amplitude of the Kolmogorov power spectrum is the strongest.

\textit{Secondly}, for each shock, we applied the same peak detection criteria, which is outlined in Appendix \ref{section:Peak Detection Algorithm}. The chosen criteria are successful in identifying the stronger peaks in most cases, though weaker peaks may remain undetected.

Finally, we should mention that the current PSDs are computed using the traditional FFT algorithm. A more contemporary method, such as wavelet analysis, could reveal a more persistent presence of these peaks. However, since our goal is to understand the reasons behind these peaks, we have reserved the use of more advanced analysis techniques for future work.

\subsection{Energetic particles and PSD peaks}
As depicted in Figure~\ref{case1_shock}, particles that become energized escape the shock front and create a foreshock region in the upstream. This holds true for other shocks as well. One of our primary investigations was to explore their role in the Power Spectral Density (PSD). In this context, we aim to determine whether energetic particles can interact with waves that have frequencies corresponding to the PSD peaks. To address this, we employ the methodology outlined in~\cite{Howard_2017}, where the solar wind interaction with waves in the lunar plasma environment was explored using data from two ARTEMIS spacecraft.

The waves in the shock upstream are seen to be predominantly right-hand polarized, as observed by \cite{Zhao_2021}. Following the equation for the Doppler-shifted frequency, one can express the observed frequency as follows:

\begin{equation}\label{eq:5}
f_{ob}=f \pm \dfrac{\bm{k} \cdot \bm{V}_{sw}}{2 \pi}
\end{equation}

Here, $\bm{k}$ represents the wave vector, and $\bm{V}_{sw}$ denotes the solar wind velocity. $f$ is the intrinsic frequency of the wave, which is nearly the same as that in the solar wind frame. We adopt a convention in which positive and negative frequencies indicate right- and left-handed polarization. The Doppler shift equations for these intrinsically right- and left-hand polarized waves are expressed by the + and - symbols in the above equation. 

\begin{figure*}[htbp!]
  \centering
    \includegraphics[width=0.45\textwidth,angle=0]{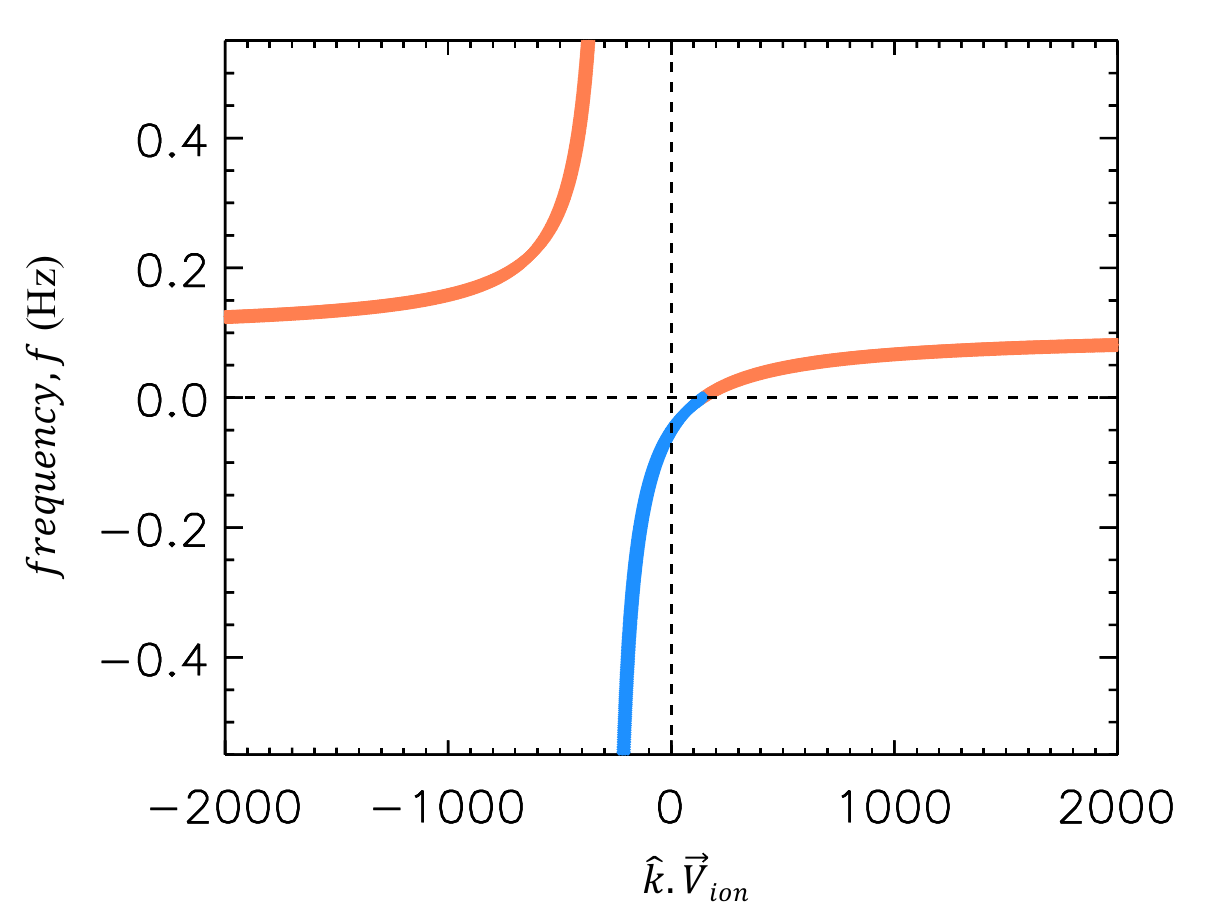}
    \includegraphics[width=0.45\textwidth,angle=0]{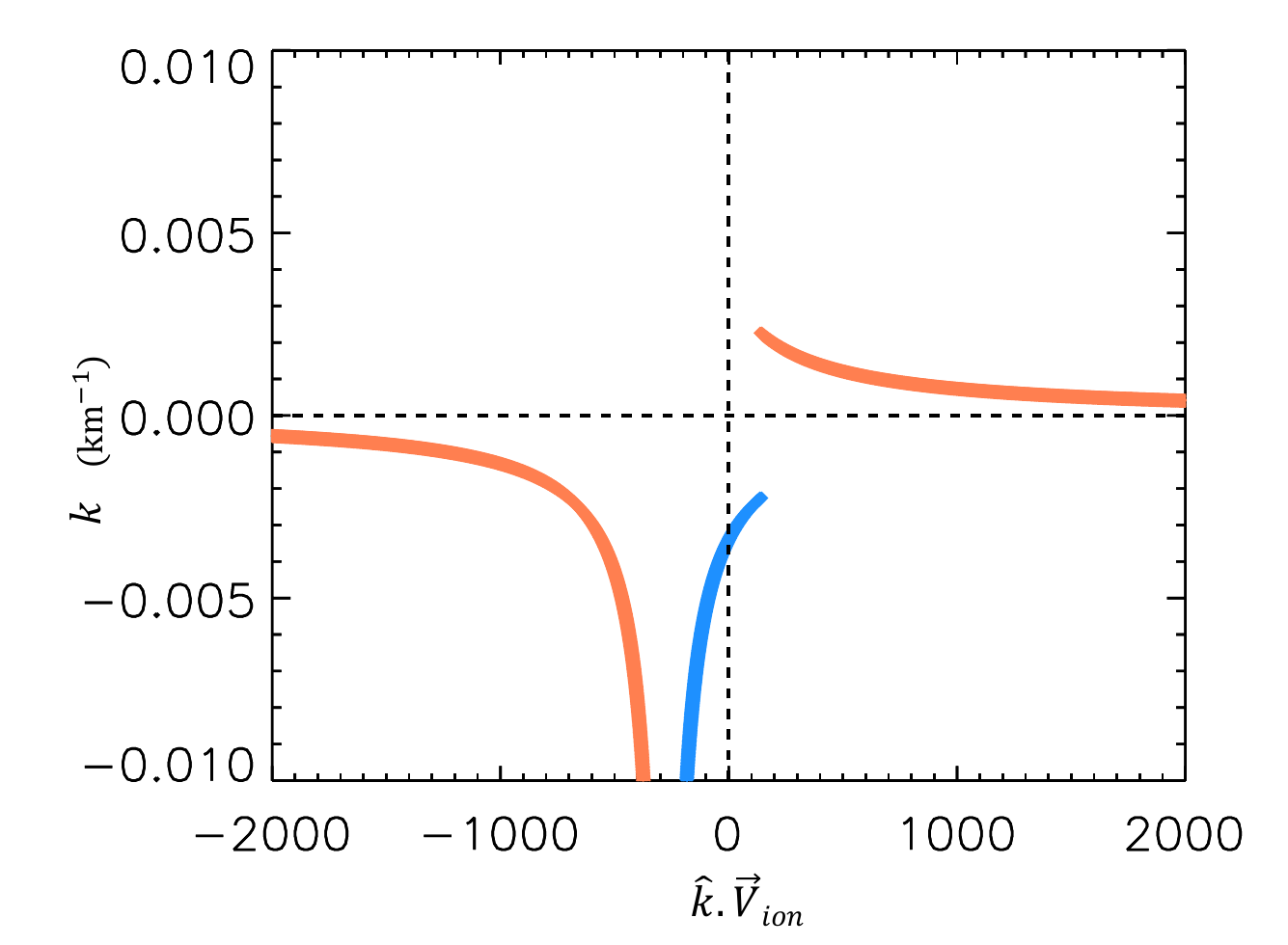}
       \caption{Intrinsic frequency ($left$) and wave number ($right$) of the wave as a function of $\bm{\hat{k}} \cdot \bm{V}_{ion}$. The orange and blue curves represent right- and left-handed  polarized waves, respectively.}
  \label{fig:dispersion}
  \end{figure*}
For ions to interact resonantly with the waves, the waves in the ion frame must be left-hand polarized, and their frequency must be the integral multiple of ion-cyclotron frequency. Considering only the first harmonic we may write

\begin{equation}\label{eq:6}
    f \mp \dfrac{\bm{k} \cdot \bm{V}_{ion}}{2 \pi}= \frac{\Omega_i}{2 \pi}
\end{equation}

Here, $\Omega_i$ represents the ion-cyclotron frequency, and $\bm{V}_{ion}$ is the velocity of the ions in the solar wind frame. 
The negative and positive symbols are used for intrinsically right and left-handed polarized waves, respectively. 
Combining Equation \ref{eq:5} \& \ref{eq:6} to solve for $f$ \& $k$ we get

\begin{equation}
    f= \dfrac{\hat{\bm{k}} \cdot \bm{V}_{sw}\frac{\Omega_i}{2 \pi}+\hat{\bm{k}}\cdot \bm{V}_{ion} f_{ob}}{\hat{\bm{k}}\cdot \bm{V}_{ion}+ \hat{\bm{k}}\cdot \bm{V}_{sw}}
\end{equation}

\begin{equation}
    k = \pm 2 \pi \dfrac{f_{ob} - \frac{\Omega_i}{2 \pi}}{\bm{\hat{k}}\cdot \bm{V}_{ion}+\bm{\hat{k}}\cdot \bm{V}_{sw}}
\end{equation}

where positive and negative symbols are used for intrinsically right- and left-handed polarized waves, respectively. For simplicity we assume the wave propagation direction ($\bm{\hat{k}}$) to be parallel to the mean magnetic field direction and is directed towards upstream. Hence positive and negative values of $k$ imply upstream and downstream propagating waves, respectively.

In Figure~\ref{fig:dispersion}, we illustrate the intrinsic frequency and wave number of the waves in relation to $\bm{\hat{k}} \cdot \bm{V}_{ion}$. The plot is generated by considering the parameters averaged over a 5-minute time window during a typical power spectral density peak. The values considered here are the local ion-cyclotron frequency $\Omega_i=0.32$ s$^{-1}$, the angle between the solar wind velocity and the mean magnetic field $\theta=43.5^\circ$, the projected solar wind velocity onto the assumed wave direction (mean magnetic field) $\hat{\bm{k}} \cdot \bm{V}_{sw} \approx 278.9$ km/s, and the observed wave frequency $f_{ob}=0.1$ Hz. Notably, when waves are intrinsically right-hand polarized, the frequency is positive and marked with a orange curve, while intrinsically left-hand polarized waves are marked with a blue curve in the figure.

Except the small sector (where $\bm{\hat{k}} \cdot \bm{V}_{ion} > 0$), the majority of the blue solution corresponds to downstream propagating (as $k$ is negative) left-hand polarized waves (as $f$ is negative). Ions to interact resonantly with these waves should propagate downstream but with speed slower than the phase speed of the wave. This sets an upper limit on the ion speed projected on the wave propagation direction ($\bm{\hat{k}} \cdot \bm{V}_{ion}$) such that the ions never overtake the waves and avoid seeing the waves as right-hand polarized. Hence, the valid solution in this case should be the negative $\bm{\hat{k}} \cdot \bm{V}_{ion}$ range as the ions should move in downstream region in contrast to $\bm{\hat{k}}$ which is upstream directed.

The orange solution corresponds to two different cases -- (a) downstream propagating (where $k$ is negative) right-hand polarized waves ($f$ is positive) and (b) upstream propagating (where $k$ is positive) right-hand polarized waves ($f$ is positive). In the former case, resonance interaction between the ions and waves can occur if ions propagate downstream at a speed faster than the phase speed of the wave. After overtaking the wave, ions perceive the wave as left-hand polarized. Hence, the solution is given by the negative $\bm{\hat{k}} \cdot \bm{V}_{ion}$ with a lower limit provided by the wave's phase speed. In the latter case, ions propagating in the upstream direction faster than the wave's phase speed can interact resonantly with the upstream propagating right-hand polarized waves. Consequently, ions with higher energies are also potential candidates to participate in the wave-particle resonance interaction, potentially leading to the observed unstable modes.

Therefore, it can be interpreted that, even though we do not find a direct signature of the interaction between the waves and high-energy protons, waves with frequencies similar to the PSD peaks are still capable of resonantly interacting with high-energy particles if they are perceived as left polarized to the protons.

\section{Summary and Discussion}\label{Section:Summary}

This study investigates the interaction between high-energy particles and solar wind turbulence in the upstream regions of six ICME-driven shocks -- five quasi-perpendicular and one quasi-parallel. All of them are standalone shocks and are not influenced by any other disturbances. The presence of enhanced upstream energetic particle fluxes served as an additional selection criterion, allowing a focused analysis of their interaction with upstream turbulence. Our key findings are summarized as follows.

\paragraph{The distinct peaks in power spectral densities indicate wave activity} Distinct peaks in the magnetic power spectral density (PSD) are intermittently observed across all shock upstreams (e.g., Figure~\ref{psd_1} (b)). These peaks serve as clear indicators of wave activities or instabilities occuring in the shock upstream. We further characterize these peaks as resulting from fluctuations in the transverse magnetic field direction. Subsequently, we explore various possibilities concerning their interaction with energetic particles generated due to ICME shocks.

Initially, we investigate whether the peaks are attributed to the alpha-proton drift or proton temperature anisotropy - common factors observed in the solar wind~\citep{Bale_2009, Hellinger_2006, Sofiane_2011}. The majority of our analyzed points fall comfortably within the instability limits (Figure~\ref{fig:thermal_check},~\ref{fig:mirror_modes}). However, a few observed points surpass the limits of proton-cyclotron or parallel firehose instability with no indication of alpha-proton drift driven instabilities.

\paragraph{Resonant interaction between Alfv\'en wave and particles}
We further explore whether the resonant streaming instability between Alfv\'en waves and protons can be responsible for the PSD peaks. Our analysis indicates that the frequencies of the PSD peaks are well-suited for resonant interaction with both the bulk solar wind protons and suprathermal protons at lower energies, reaching up to approximately $10$ keV (Figure~\ref{barplot_with_energy_case2}). We also suspect that PSD peaks due to the resonant interaction of waves and high-energy particles may not be easily observed, as the flux of high-energy particles is always much lower than that of low-energy particles. Additionally, the resonant frequency of high-energy particles would appear at even lower frequencies, where background turbulence is more pronounced.

\paragraph{Background solar wind condition in detecting PSD peaks} Among the six shocks analyzed, we found that PSD peak detectability was poor in certain cases (e.g., Shock 5 and 6). To investigate this, we examined the nature of the background solar wind conditions. Using a two-sample KS test, we concluded that the background solar wind conditions differ statistically (Figure~\ref{fig:statistics}) in the $\frac{\overline{\delta B^{2}}}{\overline{B^{2}}}$ -- $\theta$ space when PSD peaks are detected versus when they are not.  

\paragraph{It is possible for high energy particles to interact resonantly with waves} While our results suggest that the observed PSD peaks are due to interactions between solar wind particles and waves, our analysis of the relationship between frequency ($f$) and $\bm{\hat{k}} \cdot \bm{V}_{ion}$ (Figure~\ref{fig:dispersion}) has demonstrated that faster propagating suprathermal particles or more energetic particles can potentially interact resonantly with the intrinsically right-polarized upstream propagating waves, provided they propagate with a speed faster than the phase speed of these waves.

It is well known that ICME shocks accelerate particles through various processes. However, in the case of quasi-perpendicular shocks, it is rare to detect an enhancement of high-energy particles upstream, as the large angle between the shock normal and the ambient magnetic field typically prevents shock-accelerated particles from escaping into the upstream region. The observation of high-energy particles upstream of the analyzed quasi-perpendicular shocks highlights the importance of interpreting the shock-normal angle ($\theta_{BN}$) as a local and temporally variable quantity. Given the highly corrugated nature of the shock front, the local $\theta_{BN}$ continuously evolves as the shock propagates. The presence of upstream high-energy particles suggests that, at certain instances, the local angle must have transitioned to a quasi-parallel configuration, allowing particle escape into the upstream region, or processes such as shock drift acceleration have energized particles sufficiently to overcome the potential barrier of a perpendicular shock. This paper investigates whether these accelerated particles can interact with waves in the shock upstream. We utilize a simple Fourier spectrum in the present analysis to generate PSD, and our approach is based on the assumption that the wave is propagating along the mean magnetic field direction. This methodology can be further improved by employing contemporary techniques, such as wavelet analysis, or by specifically identifying the wave direction using Minimum Variance Analysis.

\section{Acknowledgments}

We thank the anonymous referee for their valuable comments, which helped improve the quality of the article. The research work at Physical Research Laboratory, Ahmedabad, was funded by the Department of Space, Government of India. SB was supported by the grant 80NSSC23K0776. SSM acknowledges support through the GSFC Heliophysics Internal Scientist Funding Model competitive work package program. All the solar wind and magnetic field data are obtained from the NASA Goddard Space Flight Center Coordinated Data Analysis Web [\url{http://cdaweb.gsfc.nasa.gov/}]. Shock parameters are obtained from the shock database [\url{http://ipshocks.fi}].

\appendix

\section{Peak Detection Algorithm}
\label{section:Peak Detection Algorithm}
 \begin{figure}
  \centering
     \includegraphics[width=0.9\textwidth,angle=0,page=1]{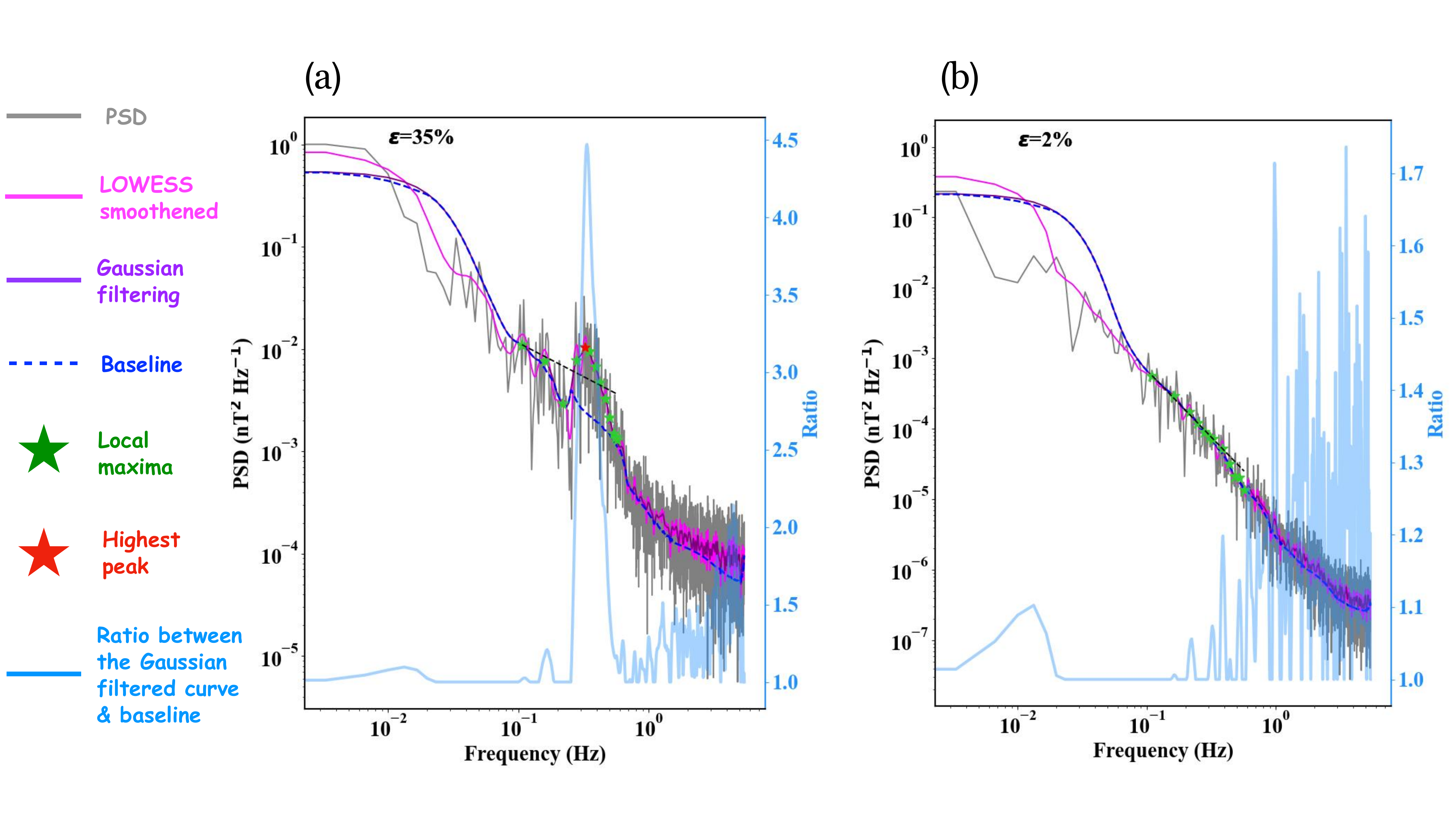}
       \caption{Examples illustrating the functionality of the peak detection algorithm. In the left PSD, a peak is identified by the algorithm as the deviation from a true power law, quantified by $\epsilon$, exceeds the threshold value of $\epsilon_{th} = 13\%$. Conversely, the PSD on the right shows no enhancement or peak, with a very small $\epsilon$ value, confirming that the power spectrum closely follows a power law.}
  \label{fig:PSD_detect}
  \end{figure}

Since multiple shocks produce PSD peaks intermittently, adopting a methodology that can automatically detect PSD peaks in all shocks is preferable and can also be used in future detection. In the following, we elaborate on the peak detection algorithm.

\begin{enumerate}[label=Step-\arabic*:]

    \item PSD from 5 minutes time window -- Each PSD is generated over a five-minute time window, as explained in Section \ref{Section:Data analysis}. Examples of PSDs are shown in Figure \ref{fig:PSD_detect}(a) and Figure \ref{fig:PSD_detect}(b) with the gray lines.

    \item PSD smoothening-I – The original PSD curves are smoothed using the LOWESS (Locally Weighted Scatterplot Smoothing) algorithm from Python's \texttt{statsmodels} library. This method effectively reduces noise in the curves while maintaining essential patterns, significantly decreasing high-frequency noise in the PSDs. LOWESS smoothened curves are shown by the magenta curves in Figure \ref{fig:PSD_detect}(a) and (b).

    \item PSD smoothening-II – Second smoothing is done by using a Gaussian filter of width, $\sigma  = 5$. In a Gaussian filter, $\sigma$ defines the standard deviation of the Gaussian kernel, which controls the degree of smoothing. A larger $\sigma$ value results in a wider and smoother filter, causing greater blurring by averaging over a broader range of neighboring data points. We have optimized this value to ensure that prominent peaks are accurately captured in the LOWESS smoothed curve. Gaussian filtered curve is shown by the violet curve in Figure \ref{fig:PSD_detect}(a) and (b).

    \item Find the PSD ‘baseline’ – The \texttt{correct\textunderscore baseline} method in  \texttt{hplc-py} is used to determine the baselines for the PSD curves, enhancing peak detection by providing clear contrast between the PSD curves and the baseline. Baselines are shown in blue in Figure \ref{fig:PSD_detect}(a) and (b). Defining a baseline helps to distinguish true peaks from background noise by establishing a reference level against which signal variations can be measured. This ensures that only significant deviations from the baseline are identified as peaks, reducing false detections caused by fluctuations in the data.

    \item Power law fitting – PSDs exhibit significant noise at higher frequencies. Additionally, a preliminary visual inspection reveals that the majority of PSD peaks are concentrated within the frequency range of 0.02–0.6 Hz. In this chosen frequency range, we perform a power law fit (using \texttt{curve\textunderscore fit} function in python) on the PSD values at the local maxima (shown by green stars in Figure \ref{fig:PSD_detect}) of the LOWESS-smoothed power spectrum. Fitted power law is shown by the black dashed line. From this fitting, we estimate the standard error of the power law slope, which indicates the quality of the fit. A smaller standard error signifies high confidence in the power law fit, while a larger standard error suggests that the slope is less precisely determined. Using the standard error, we compute a parameter, $\epsilon$, defined as:
    \begin{equation*}
        \epsilon = \frac{\text{Standard error of the power law slope}}{\text{Fitted power law slope}}\times 100 \%
    \end{equation*}
    This parameter quantifies the error relative to the power law slope. Smaller $\epsilon$ values indicate that the data points are well-fitted by a power law, resulting in minimal slope error. Conversely, when the data points (green stars) deviate from a power law, typically due to a peak in the power spectrum, $\epsilon$ exhibits larger values. We define a threshold value, $\epsilon_{th}$, such that a peak is identified only when $\epsilon$ exceeds this threshold. In this study, we use $\epsilon_{th} = 13\%$ to detect the stronger peaks in the power spectrum.  $\epsilon_{th}$ is a user-defined parameter where adjusting its value allows for the detection of peaks with different strengths. 
        
    \item Peak frequency -- To determine the frequency corresponding to the PSD peak, we calculate the ratio between the Gaussian filtered PSD and the baseline. This quantity, defined as $\mathcal{R}$, determines how much above the baseline the peak is standing (shown by the light blue curves in Figure \ref{fig:PSD_detect}(a) and (b) with values on right Y-axis). We then consider the peak frequency as the frequency where $\mathcal{R}$ has its maximum value. For example, the PSD peak is shown by a red star in Figure \ref{fig:PSD_detect}(a). The frequency range where the PSD exhibits enhancement is determined by fitting a Gaussian to the ratio curve, with the full-width at half maximum (FWHM) of the fitted Gaussian representing the frequency range.

\end{enumerate}

Although we have visually verified that most of the detected peaks meet our satisfaction, we would also like to highlight that, many times, automatic detection of PSD peaks in this manner may contribute to erroneous identification for the following reasons: (a) The PSD peaks that we find often lie in the region where the spectral break occurs, marking the transition from the inertial to the kinetic range of turbulence. This break could be mistakenly identified as a peak, leading to an overestimation of ``peaky'' intervals. (b) Energetic particles may excite a broad spectrum of waves, enhancing fluctuations across a wide range of wave numbers and creating a \textit{power-law-like} PSD within the analyzed range. This effect could lead to an underestimation of intervals exhibiting wave-particle interactions. Indeed, these challenges make it difficult to jointly highlight energetic particles and spectral peaks in the PSD. 

\section{Shock details}
\label{section:SD}
\begin{figure}
  \centering
      \includegraphics[width=0.9\textwidth,angle=0,page=3]{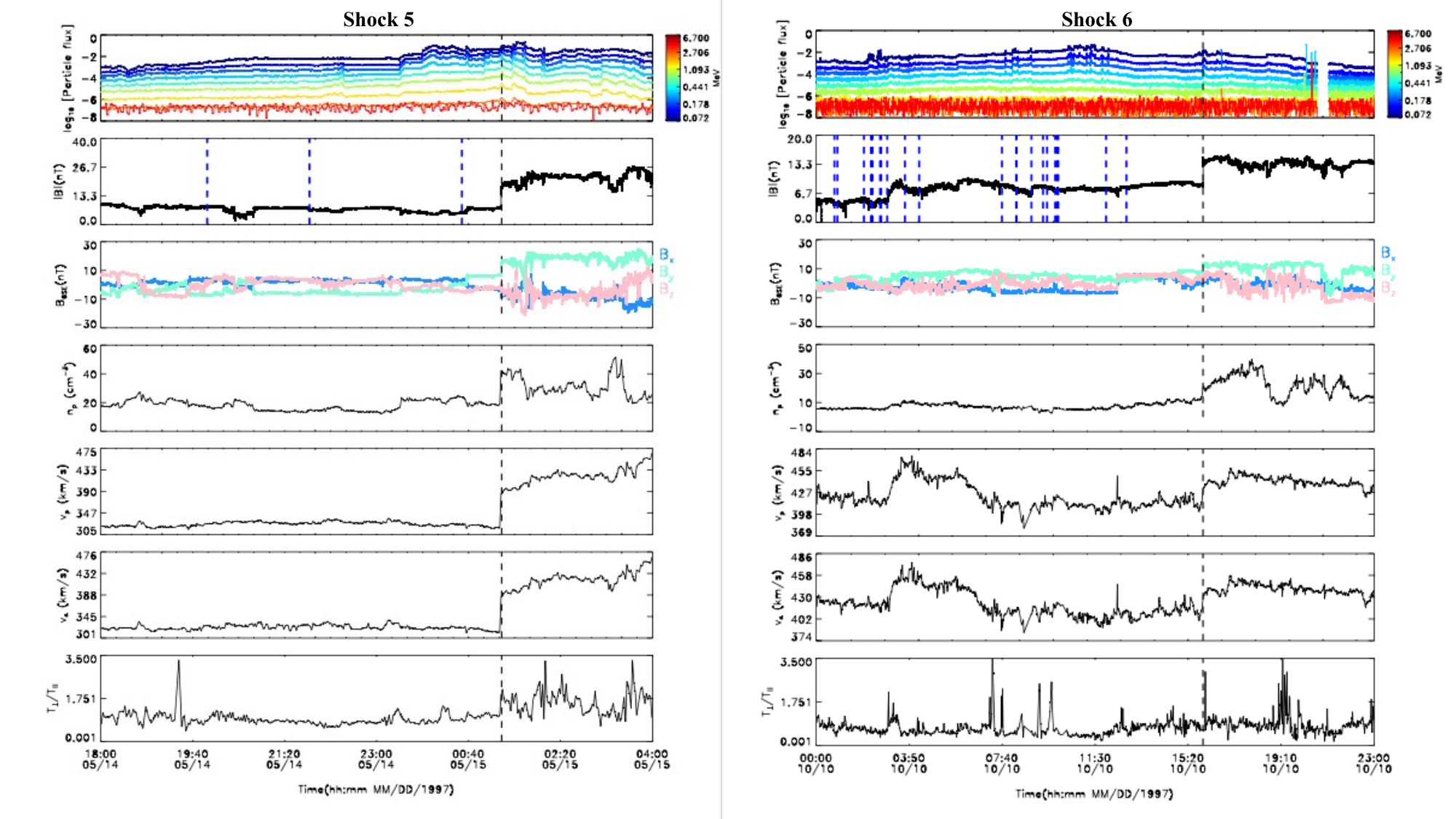}
       \caption{Plasma parameters for the upstream and downstream regions of all six shocks analyzed in this study are shown. Each plot is labeled with the corresponding shock number as listed in Table \ref{tab:table_1}. The vertical black dashed line in each plot denotes the shock arrival time. For each shock, the top panel displays the particle flux of various energetic particles (from $72$ keV to $6.7$ MeV) near the shock occurrence, followed by parameters such as the magnitude of the magnetic field ($|B|$), the three components of the magnetic field in the GSE coordinate system ($B_x$, $B_y$, $B_z$), solar wind particles density ($n_p$), bulk proton velocity ($V_p$), bulk $\alpha$ particle velocity ($V_{\alpha}$), and the evolution of temperature anisotropy ($T_{\perp}/T_{\parallel}$) around the time of the shock arrival. The blue dashed lines in the total magnetic field panel mark the time intervals of PSD peaks.}
  \label{fig:peaks_in_all_shocks}
  \end{figure}

A list of the analyzed shocks is provided in Table~\ref{tab:table_1}. In situ measurements for all these selected shocks are depicted in Figure~\ref{fig:peaks_in_all_shocks}. In each plot, the blue dashed vertical lines highlight the intervals when we identify the PSD peaks.

 \begin{figure}
  \centering
      \includegraphics[width=0.9\textwidth,angle=0,page=1]{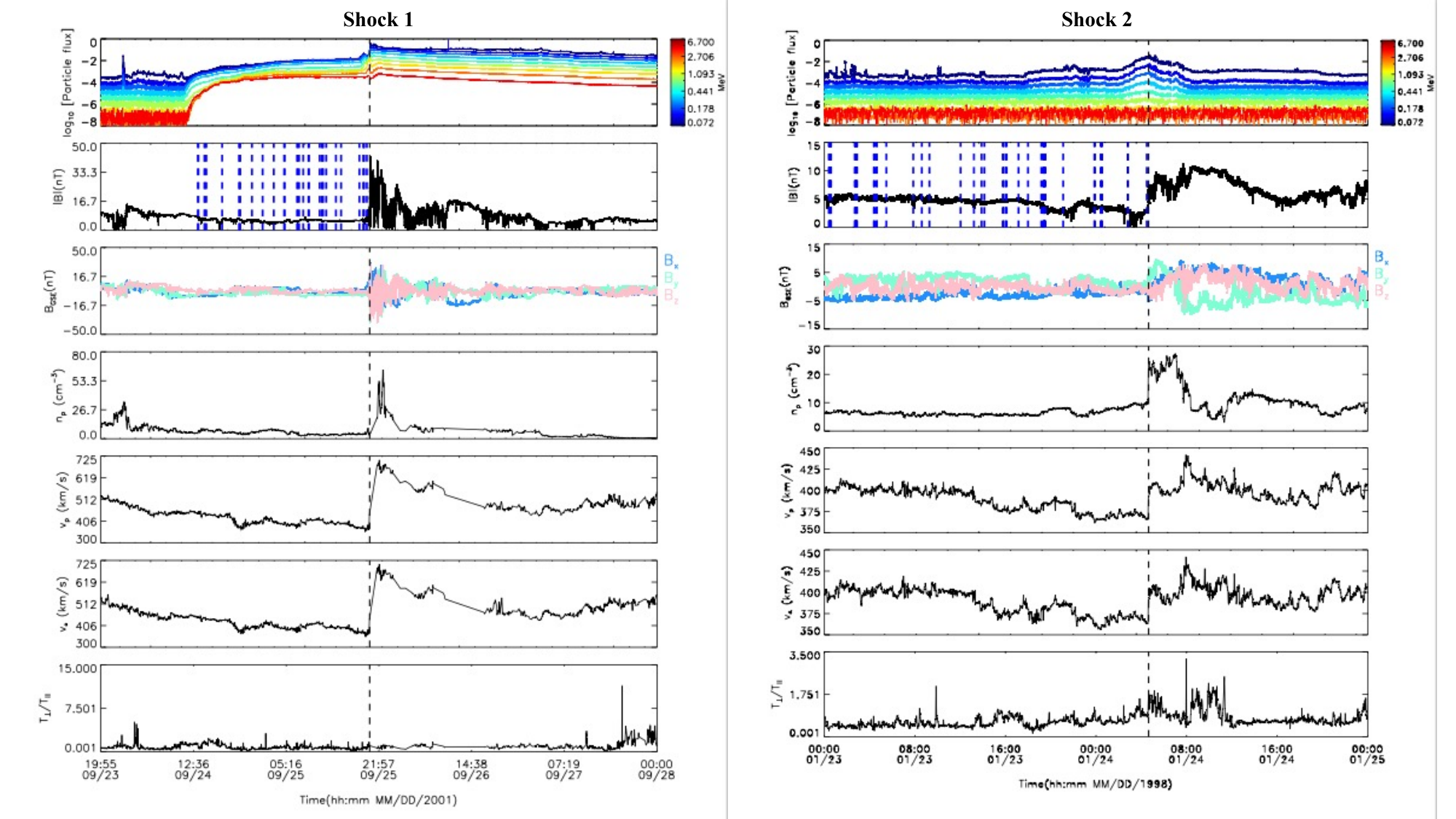}
      \includegraphics[width=0.9\textwidth,angle=0,page=2]{figures_new/all_peaks_auto_nw.pdf}
  \end{figure}

\bibliography{ms}{}
\bibliographystyle{aasjournal}

\end{document}